\documentclass[twoside,10pt]{article}

\usepackage{1}

\title{\Large \bf Effects of the next-to-leading order terms in the chiral SU(3) Lagrangian  on the strangeness -1 s-wave meson-baryon interactions }
\author{V.K. Magas$^{1,2}$,  A. Feijoo$^1$, A. Ramos$^{1,2}$}
\date{}
\begin{document}
\maketitle

\begin{center}
\vspace*{-0.3cm}
{\it  $^1$~Dept. d'Estructura i Constituents de la Mat\`eria, Universitat de Barcelona, Mart\'i Franqu\`es 1, E08028 Barcelona, Spain}\\
{\it  $^2$~Institut de Ci\'encies del Cosmos, Universitat de Barcelona,\\
 Mart\'i Franqu\`es 1, E08028 Barcelona, Spain}%\\
% e-mail: }
\end{center}

\vspace{0.3cm}

\begin{center}
{\bf Abstract}\\
\medskip
\parbox[t]{10cm}{\footnotesize
The meson-baryon interactions in s-wave in the strangeness S=-1 sector are studied using a chiral unitarity approach based on the next-to-leading order chiral SU(3) Lagrangian. The model is fitted to the large set of experimental data in different two-body channels. Particular attention is paid to the $\Xi$ hyperon production reaction, 
$\bar{K} N \rightarrow K \Xi$, where the effect of the next-to-leading order terms in the Lagrangian play a crucial role, since the cross section of this reaction at tree level is zero. 
 }
\end{center}

%\section{Introduction} \label{s1}
%\setlength{\parskip}{1mm}
It is well known that the proper theory of strong interactions -  Quantum Chromodynamics (QCD) - is not suitable to study low energy hadron dynamics. For such studies effective theories should be used, and SU(3) Chiral Perturbation Theory  ($\chi$PT) is a classical example. This theory is based on an effective Lagrangian with hadron degrees of freedom, which respects the symmetries of QCD, in particular the chiral symmetry $SU(3)_R\times SU(3)_L$. $\chi$PT has many succesful applications, however it fails to describe the hadron dynamics in the vicinity of dynamically generated resonances. A good example of such situation is the kaon-nucleon interaction at low momenta, where the perturbation scheme is violated by the presence of the  $\Lambda(1405)$ resonance, located only $27$ Mev below the $KN$ threshold. In this case the use of some non-perturbative techniques is mandatory. In particular such a situation can be successfully studied within a unitary extension of Chiral Perturbation Theory (U$\chi$PT), originally proposed in \cite{ref5}, where the unitarization is implemented in coupled channels. 

The $\Lambda(1405)$ resonance does not only give a reason to use U$\chi$PT theory, but also provides a good test of the predictive power of this approach. The point is that the $\Lambda(1405)$ is a dynamically generated resonance. This was predicted  for the first time  in 1977, see Ref. \cite{L1405}, and later detailed calculations performed in the framework of U$\chi$PT have shown that $\Lambda(1405)$ is actually a superposition of two close dynamically generated states: one at lower energy $\approx 1390$ MeV with larger width $\approx 130$ MeV, which couples most strongly to $\Sigma \pi$ channels; and the other one at higher energy $\approx 1420$ MeV and with a much narrower width $\approx 30$ MeV, which couples most strongly to $\bar{K} N$ channels. Thus, the experimental shape of the $\Lambda(1405)$ resonance depends on the details of the given experiment, namely on the relative weight of the $\Sigma \pi$ and $\bar{K} N$ channels in the given reaction. This rather nontrivial prediction has been finally confirmed experimentally, see Ref. \cite{PRL} for more details. 

The U$\chi$PT method consists in solving the Lippmann-Schwinger equations in coupled channels, which is reduced to a system of  algebraic equations \cite{ref1}:
\begin{equation}
T_{ij} =V_{ij}+V_{il} G_l T_{lj}\,,
\label{LS}
\end{equation} 
where $T_{ij}$ is the scattering amplitude for the transition from channel "i" to channel "j"; the subscripts %$i, j, l$ 
run over all the possible channels. In particular, for the meson-baryon interaction in the $S=-1$ sector, which is of prime interest for us,  there are the following 10 channels: $K^- p$, $\bar{K}^0 n$, $\pi^0 \Lambda$, $\pi^0 \Sigma^0$, $\pi^+ \Sigma^-$, $\pi^- \Sigma^+$, $\eta \Lambda$, 
$\eta\Sigma^0$, $K^+ \Xi^-$, $K^0 \Xi^0$. 

In our study we calculate the loop function, $G_l$, using a dimensional regularization scheme: 
$$
G_l=\frac{2M_l}{(4\pi)^2}\left\{a_l+\ln\frac{M_l^2}{\mu^2}+\frac{m_l^2-M_l^2+s}{2s}\ln\frac{m_l^2}{M_l^2}+\right.
$$
\begin{equation}
\left.+\frac{q_{cm}}{\sqrt{s}}\ln\left[\frac{(s+2\sqrt{s}q_{cm})^2-(M_l^2-m_l^2)^2}{(s-2\sqrt{s}q_{cm})^2-(M_l^2-m_l^2)^2}\right]\right\}\,,
\end{equation}
where $M_l$ and $m_l$ are the baryon and meson masses of the "$l$" channel correspondingly, and $a_l$ are the so called subtraction constants, which are used as free parameters and fitted to the experimental data. Taking into account the isospin symmetry there are only 6 independent subtraction constants. See Ref. \cite{ref1} for more details.

The function $V_{ij}$ is the interaction kernel for $(i,j)$ channels, which is calculated from the chiral Lagrangian up to the corresponding order in momentum over baryon mass. For the meson-baryon interaction, the lowest order term in momentum, i.e. leading order (LO) term, is the so called Weimberg-Tomozawa (WT) term: 
\begin{equation}
V^{WT}_{ij}=-C_{ij}\frac{1}{4f^2}(k^0+k'^0)\,, 
\label{WT}
\end{equation}
which depends only on one parameter - the pion decay constant $f$. $C_{ij}$ is a matrix of coefficients; $k^\mu$ and  $k'^\mu$ are the four-momenta for the incoming and outgoing mesons in the process. The pion decay constant is well known experimentally, $f_{exp}=93.4$ MeV, however in LO U$\chi$PT calculations this parameters is usually taken to be $f=1.15-1.2 f_{exp}$, in order to partly simulate the effect of the higher order corrections.

The interaction kernel up to next-to-leading order (NLO) is also known:
\begin{equation}
V^{NLO}_{ij}=V^{WT}_{ij}+\frac{1}{f^2}\left(D_{ij}-2(k_{\mu}k'^{\mu})L_{ij}\right)\sqrt{\frac{M_i+E_i}{2M_i}}\sqrt{\frac{M_j+E_j}{2M_j}}
\label{V_NLO}
\end{equation} 
where  $D_{ij}$ and $L_{ij}$ are the coefficient matrixes, which depend on the new parameters: $b_0$, $b_D$, $b_F$, $d_1$, $d_2$, $d_3$, $d_4$ (see \cite{ref2,ref3,ref4,ref25,ref22} for more details).
However only very recently it started to be used in real calculations and data fitting \cite{ref2,ref3,ref4,ref25,ref22}. The reason is rather straightforward - NLO terms in chiral Lagrangian depend on 7 new parameters, which were not known, and thus the predictive power of the NLO  U$\chi$PT calculations was rather questionable. 
  
Thanks to great experimental advances of the last years, like for example CLAS photoproduction experiments \cite{CLAS},  we have accumulated a sufficient amount of a good quality data to attempt to fit these new parameters. Also due to the large amount of theoretical studies based on the WT interaction we know where this approach fails to describe the data. In particular, in our study we concentrate on the $\Xi$ hyperon production reactions: $K^- p \rightarrow K^+ \Xi^-\,, K^0 \Xi^0$, where the effect of the NLO terms in the Lagrangian play crucial role, since the cross sections are zero at tree level.  These reactions are also particularly interesting, because they were not considered in the works of the other groups \cite{ref2,ref3,ref4,ref22}.

% To include figures
\begin{figure}[htb]
\begin{center}
\includegraphics[width=10.5cm]{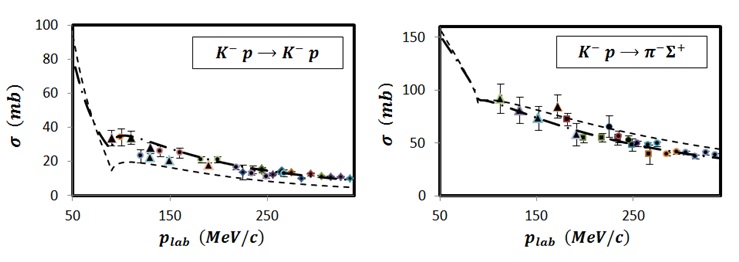}
\end{center}
\vspace{-0.7cm} 
\caption{The total cross section of the $K^- p$ scattering to the indicated channels. WT fit is presented by  the dashed line; NLO fit - by the dot-dashed line. Experimental data are from \cite{ref7,ref8,ref9,ref10}. } 
\vspace{-0.2cm}
\label{fig1}
\end{figure}

\begin{figure}[h]
\begin{center}
\includegraphics[width=10.5cm]{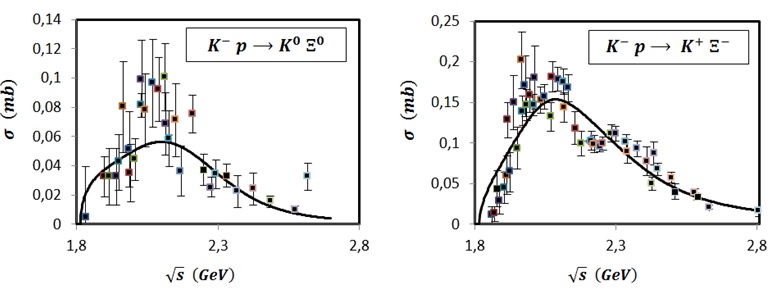}
\end{center}
\vspace{-0.7cm} 
\caption{The total cross section of the $K^- p$ scattering to the indicated channels.  The solid line represents results of NLO fit (the WT interaction is 0). Experimental data are from \cite{ref13,ref14,ref15,ref16,ref17,ref18,ref19}. } 
\vspace{-0.2cm}
\label{fig2}
\end{figure}

Having calculated the T-matrix, by solving the system of equations (\ref{LS}), we can then calculate the corresponding cross section for the $i \rightarrow f$ reaction in the following way: 
$$
\sigma_{if} =\frac{1}{4\pi}\frac{M_i M_f}{s}\frac{k_f}{k_i}|T_{if}|^2\,.
$$
To simulate the available experimental data we calculate the transitions from the $K^- p$ initial state to different final states and study the $\sqrt{s}$ dependence of the corresponding cross sections; some examples are presented in Figs. \ref{fig1} and \ref{fig2}, and three branching ratios are compared to the experimental data in Table \ref{tab1}. We perform 7 and 14 parameter fits:  the pion decay constant and 6 subtraction constants in the case of the WT interaction kernel (WT fit), and 7 additional NLO parameters for the NLO kernel (NLO fit).     

%To include Tables
\begin{table}[htb] \noindent
\caption{This table shows the branching ratios at threshold for the best $\chi^2$ fit at tree level (WT) and for NLO calculations, to be compared with experimental values. }
\vskip3mm\tabcolsep5pt
\begin{center}
{\footnotesize\begin{tabular}{|c c c c |}
 \hline 
 \multicolumn{1}{|c} {} & \multicolumn{1}{|c}{$\gamma=$}& \multicolumn{1}{|c}{$R_n=$}& \multicolumn{1}{|c|}{$R_c=$}\\%
 \multicolumn{1}{|c} {} & \multicolumn{1}{|c}{$\frac{\Gamma(K^- p \rightarrow \pi^+ \Sigma^-)}{\Gamma(K^- p \rightarrow \pi^- \Sigma^+)}$}& 
 \multicolumn{1}{|c}{$\frac{\Gamma(K^- p \rightarrow \pi^0 \Lambda)}{\Gamma(K^- p \rightarrow neutral\, states)}$}& 
 \multicolumn{1}{|c|}{$\frac{\Gamma(K^- p \rightarrow \pi^+ \Sigma^-,\pi^- \Sigma^+ )}{\Gamma(K^- p \rightarrow inelastic\, channels)}$}\\%
\hline%
WT &	2.25	& 0.196	& 0.636  \\%
NLO &	2.36 &	0.197 &	0.659  \\%
Exp. &	$2.36\pm 0.04$ &	$0.189\pm 0.015$ & $0.664\pm 0.011$ \\
\hline
\end{tabular}}
\label{tab1}
\end{center}
\vspace{-0.8cm} 
\end{table}

Looking at Figs. \ref{fig1}, \ref{fig2} and Table \ref{tab1} we can conclude that the inclusion of the NLO terms into the interaction kernel improves the agreement with data. This has also been shown in more detail in Refs. \cite{ref2,ref3,ref4,ref25,ref22}.  

Fig. \ref{fig2} shows the $\Xi$ hyperon production reactions, which are the key point of this work, as well as of the previous study \cite{ref25}.  Note that these channels are extremely sensitive to the NLO corrections and therefore may play a crucial role in determining the NLO parameters. However, before coming up with the final result of the fit we have to make sure that we have taken into account all the physical processes, significant for these reactions. We can see that although the overall agreement is 
%reasonable, 
quite good,
still there are some resonance-like shapes in the experimental data, which are not reflected in our smooth curves.  In our opinion, which is based on the phenomenological study \cite{ref12}, this indicates the necessity to take into consideration the  $\bar{K} N \rightarrow Y \rightarrow K \Xi$ reactions, where $Y$ stands for some high spin resonances, which couple to these channels. Based on our results, Fig. \ref{fig2}, it seems that the $\Sigma(2030)$ and the $\Sigma(2250)$ resonances would be good candidates for "$Y$", and this observation coincides with the finding of \cite{ref12}. These resonances have spins ${5/2}^-$ and ${7/2}^+$  respectively, and therefore require a special treatment, analogous to that performed in \cite{ref23,ref24}. This work is in progress now.

The final goal of our study is to find trustable restrictions on the 7 NLO parameters of the chiral Lagrangian. We would like to stress that, technically, to change in the calculations the WT interaction, eq. (\ref{WT}), 
to the NLO interaction, eq. (\ref{V_NLO}), is rather straightforward. The problem comes from the fact that the 7 new parameters of the NLO interaction are not well controlled at the moment. Once stable values 
for these parameters are obtained, all the groups doing simulations based on the chiral Lagrangian will be able to increase the accuracy of their calculations to the next order with a rather little effort.

{\bf Acknowledgments.} \ \ This work is supported  by the European Community - Research Infrastructure Integrating Activity Study of Strongly Interacting Matter (HadronPhysics3, Grant Agreement Nr 283286) under the 7th Framework Programme, by the contract\\ FIS2011-24154 from MICINN (Spain), and by the Ge\-ne\-ra\-li\-tat de Catalunya, contract 2009SGR-1289.

\end{document}